# MemVanta: Memory-Efficient CPU Inference for Quantized GGUF Language Models

Saurav Singla Aarav Singla Advik Gupta Parnika Gupta

Abstract. Deploying large language models on commodity CPUs is often constrained less by arithmetic throughput than by the amount of resident memory that can be sustained under a fixed system budget. MemVanta is an experimental C++20 inference runtime for quantized GGUF language models that treats memory residency as a first-class systems objective. The runtime combines memory-mapped model access, bounded tensor caching, quantized execution paths, paged key-value state, and explicit memory-pressure validation. We evaluate MemVanta against a pinned llama.cpp baseline using the exact same OpenLLaMA 7B v2 Q4_0 GGUF on CPU-only workloads. Across five measured runs after warm-up, MemVanta records 3,983,412 KiB peak resident set size compared with 7,586,960 KiB for llama.cpp, a 47.50% reduction (approximately 3.44 GiB). This memory saving is not free: prompt processing is 3.15 ± 0.04 tok/s versus 45.82 ± 0.96 tok/s, and token generation is 1.52 ± 0.02 tok/s versus 7.76 ± 0.09 tok/s. A separate Linux cgroup-v2 experiment with swap disabled shows MemVanta completing at a 3584 MiB tested memory ceiling where llama.cpp is OOM-killed; llama.cpp's lowest successful tested ceiling is 3840 MiB. We therefore frame MemVanta as a memory-efficiency prototype rather than a throughput competitor. The paper details architecture, memory accounting, quantized kernels, cache design, experimental controls, failure modes, reproducibility, and the engineering trade-offs required to move toward practical edge and low-RAM CPU inference.

## 1. Introduction

Local inference has become an important deployment mode for language models because it can reduce data movement, support offline operation, simplify privacy boundaries, and make experimentation possible without dedicated accelerators. Yet the phrase “runs locally” hides a systems constraint that is especially severe on commodity CPUs: a model may fit on storage and still fail to execute within the resident-memory budget available to the process. Model weights, dequantization buffers, intermediate activations, tokenization state, thread-local scratch space, key-value cache, runtime metadata, allocator fragmentation, and operating-system page residency all contribute to the working set. In low-RAM virtual machines, mini PCs, embedded x86 systems, and shared servers, this working set can determine whether inference starts at all.

Most optimized inference engines reasonably prioritize throughput and latency. MemVanta explores a different point in the design space: what changes when the runtime treats peak resident memory as the primary optimization target and reports throughput as an explicit cost? The objective is not to claim that memory minimization is universally preferable. Rather, it is to characterize a deployment regime in which the feasibility boundary is defined by RAM. This distinction motivates both the architecture and the evaluation protocol.

The central empirical result is intentionally narrow. On the repository's repeated OpenLLaMA 7B v2 Q4_0 A/B workload, with the exact same GGUF, four CPU threads, pp512/tg128, context 768, batch 32, F16 KV, one warm-up, and five measured repetitions, MemVanta reduces measured peak RSS by 47.50% relative to the pinned llama.cpp comparison. The same experiment also shows a large throughput disadvantage. We treat both observations as inseparable parts of the result.

## 1.1 Contributions and claim boundaries

This work makes four contributions. First, it presents a compact CPU inference architecture designed around bounded residency rather than maximum throughput. Second, it defines a reproducible evaluation protocol that separates repeated peak-RSS measurement from hard memory-pressure execution. Third, it reports a same-model, same-GGUF comparison against a pinned llama.cpp baseline and retains raw evidence rather than only aggregate claims. Fourth, it analyzes the mechanisms and limitations behind the observed memory-throughput trade-off, including mmap behavior, cache bounds, KV representation, quantized kernels, batching, and page-fault sensitivity.

The paper does not claim a universal memory scaling law, a universal advantage over llama.cpp, or a throughput improvement. It does not infer minimum physical-RAM requirements from a finite cgroup sweep. It does not treat a GitHub-hosted benchmark as independent third-party replication. These boundaries are important because memory measurements are highly sensitive to model family, quantization, kernel implementation, context length, allocator behavior, kernel page cache, host configuration, and benchmark sequence.

# 2. Problem Formulation

Let a quantized language model contain weight tensors W, persistent runtime metadata M, key-value state K(t) at decoding step t, transient activation/scratch state A(t), and runtime cache C(t). A useful operational objective for constrained inference is not merely to minimize the serialized model size |W|, but to minimize the maximum resident working set R(t) observed over an inference episode subject to functional correctness. We write the systems objective as min max_t R(t), where R(t) includes resident mapped pages, anonymous allocations, thread stacks, cache state, and other process-resident memory visible to the operating system.

Throughput introduces a competing objective. Techniques that reduce memory can increase page faults, repeated tensor decoding, cache misses, synchronization, and memory traffic. The deployment problem is therefore multi-objective: maximize feasibility under a memory budget B while maintaining acceptable prompt-processing throughput P and token-generation throughput G. For a fixed workload, a configuration is feasible when max_t R(t) ≤ B and the inference process completes without OOM termination or runtime failure.

## 2.1 Why serialized model size is insufficient

A Q4_0 GGUF file provides a compact on-disk representation, but the runtime's resident footprint is not identical to file size. Memory mapping can defer residency, yet accessed pages become resident according to kernel policy. Quantized matrix operations may allocate decoded blocks or scratch buffers. KV state grows with sequence length. Parallel execution can multiply temporary state. A runtime that eagerly materializes or retains tensors may therefore consume substantially more resident memory than a runtime that streams or bounds these structures.

This distinction also explains why two runtimes executing the same GGUF can show different RSS. The difference need not imply that one stores fewer model parameters; it may reflect different choices about when pages are touched, what is copied, how long temporary state survives, how KV memory is represented, and how aggressively kernels trade additional memory for speed.

## 2.2 Measurement model

We distinguish three quantities: serialized model bytes, peak process RSS during an unconstrained run, and the lowest successful tested cgroup ceiling in a predefined sweep. These quantities answer different questions. File size describes storage. Peak RSS approximates the largest resident process footprint observed under the workload. A cgroup ceiling tests whether the runtime can complete when the kernel enforces a hard memory budget. The last quantity is a tested feasibility point, not a mathematical minimum.

# 3. MemVanta Architecture

MemVanta is implemented as a C++20 runtime with a small set of components whose interfaces are visible in the repository build graph: memory-mapped file access, tensor storage, bounded LRU caching, prefetching, runtime orchestration, matrix multiplication, quantization and quantized kernels, GGUF parsing, worker-pool parallelism, Llama-style model execution, benchmarking, profiling, auto-tuning, evaluation, and tokenizer validation. The architecture is deliberately modular so that memory-sensitive choices can be isolated and measured.

## 3.1 Memory-mapped model access

The runtime maps model data instead of unconditionally copying the entire serialized model into anonymous heap memory. Mapping alone does not guarantee a small RSS: pages become resident when touched, and the kernel can retain them in memory. The design benefit is control. Tensor access can be demand-driven, and the runtime can avoid a second full-size copy of the weight file. This is particularly useful when the model is larger than the desired anonymous-memory budget but the workload accesses weights in a predictable layer-by-layer sequence.

The operating system becomes part of the effective memory hierarchy. Sequential layer traversal benefits from read-ahead, while random access can create page-fault amplification. Consequently, MemVanta pairs mmap access with bounded caching and prefetch decisions rather than assuming mapping is automatically efficient.

## 3.2 Tensor store and bounded cache

The tensor store resolves GGUF metadata to tensor views and exposes data to execution kernels. A bounded LRU cache limits the amount of derived or repeatedly used tensor state retained across operations. The policy is intentionally conservative: retaining more transformed state can improve throughput, but it moves the runtime away from the memory-feasibility objective. Eviction therefore acts as a pressure valve that caps long-lived auxiliary residency.

The cache design creates a classic recomputation-versus-residency trade-off. When a tensor representation is evicted, later reuse may require re-reading pages or reconstructing an execution-friendly form. The measured throughput gap against llama.cpp is consistent with a design that gives up aggressive reuse and optimized kernels in exchange for a smaller working set. This is a feature of the current research question, not evidence that throughput is unimportant.

## 3.3 Prefetching under a memory objective

Prefetching can hide I/O and page-fault latency, but uncontrolled prefetching can inflate RSS by making future pages resident earlier than necessary. MemVanta therefore treats prefetch distance as a bounded scheduling parameter. The ideal distance depends on storage latency, kernel execution time, layer size, available RAM, and kernel page-reclamation behavior. A memory-first runtime should prefetch enough to overlap useful work but not so far ahead that prefetched pages coexist with large transient buffers and KV growth.

## 3.4 Worker pool and thread-local state

CPU parallelism is another source of hidden memory multiplication. Per-thread scratch buffers, reduction state, dequantization blocks, and stack growth can increase RSS with thread count. The published 7B experiments use four threads, which is therefore part of the experimental contract. Scaling thread count is future work and should be evaluated as a joint throughput-memory curve rather than as an isolated speed parameter.

# 4. Quantized Execution

MemVanta supports Q4_0, Q6_K, Q8_0, F16, and F32 execution paths. Quantization reduces serialized weight storage and memory bandwidth, but a runtime can lose part of this benefit if kernels repeatedly expand large regions into higher-precision buffers. The implementation therefore emphasizes blockwise quantized computation, where compact blocks are decoded close to the arithmetic operation and temporary state is kept small.

## 4.1 Q4_0 path

Q4_0 is central to the headline 7B experiment. In a block-quantized representation, groups of low-bit values share scale metadata. A memory-efficient kernel should consume these blocks without materializing an entire full-precision matrix. The practical design challenge is to balance compactness with vectorization. Small decode buffers preserve memory but can increase instruction overhead and reduce arithmetic intensity; large decoded tiles improve reuse but increase transient residency.

MemVanta includes AVX2/FMA-oriented quantized kernels. The current implementation is best understood as a correctness- and memory-oriented baseline rather than a fully mature CPU kernel library. llama.cpp has years of architecture-specific optimization, hand-tuned kernels, scheduler refinements, and broad hardware coverage. The large prompt-processing gap observed in our comparison is therefore expected and is reported explicitly.

## 4.2 Mixed representation and temporary buffers

Memory accounting must include not only persistent quantized weights but also temporary representations created by matrix operations. For each operator, the relevant quantity is the maximum live set: input activation, output activation, quantized weight blocks, decoded tiles, accumulation buffers, thread-local partitions, and any scheduler metadata that overlap in time. Reducing peak RSS often requires shortening lifetimes as much as shrinking individual allocations.

## 4.3 Batched prefill versus autoregressive decode

Prompt processing and token generation stress the runtime differently. Prefill processes many prompt tokens and can benefit substantially from batching, cache-friendly matrix operations, and optimized parallelism. Decode repeatedly performs small-step autoregressive work and becomes sensitive to KV access, memory latency, and per-token scheduling overhead. This difference is visible in the results: MemVanta's relative throughput deficit is larger during prompt processing than during token generation.

## 4.4 Numerical validation

A memory-saving optimization is useful only if it preserves executable model behavior. MemVanta therefore separates systems benchmarks from numerical and tokenizer validation. The repository contains tokenizer fixtures for GPT-2/SmolLM-style and Llama/SentencePiece-style tokenization, model execution smoke tests, and evaluation tooling. For a mature release, this validation should expand to layer-level reference comparisons, logit tolerances across quantization formats, deterministic prompt suites, and task-level quality metrics.

A key methodological principle is that lower RSS alone cannot establish a superior inference engine. A valid comparison requires the same model artifact, compatible tokenization, equivalent context and KV settings, comparable prompt/generation lengths, and evidence that both runtimes complete meaningful inference. The benchmark protocol is designed around this principle.

# 5. KV Cache and Sequence-State Memory

Autoregressive inference maintains key and value tensors for prior tokens. KV memory grows with sequence length and can dominate incremental memory for long contexts. MemVanta provides paged F32, F16, and Q8 KV options, allowing the runtime to trade precision, memory, and compute overhead. The published headline comparison uses F16 KV and context 768, so the 47.50% RSS result should not be extrapolated to longer contexts or different KV formats without measurement.

## 5.1 Paged allocation

A paged KV design avoids requiring one monolithic contiguous allocation for the maximum theoretical context. Pages can be allocated as sequence state grows, which improves control over commitment and can reduce fragmentation. Paging also provides a natural unit for future eviction, compression, or offload strategies, although those mechanisms introduce their own latency and correctness considerations.

## 5.2 KV precision

F32 offers a straightforward reference representation but doubles storage relative to F16. Q8 can reduce KV storage further, but quantizing dynamic state may affect accuracy and introduces encode/decode overhead. Because KV is accessed repeatedly during attention, its representation can affect both memory bandwidth and token-generation latency. A complete Pareto analysis should therefore report quality, RSS, and throughput for each KV mode across multiple context lengths.

## 5.3 Context-length scaling

The current 768-token context is intentionally modest. For edge deployments, context length is often a configurable resource budget rather than a fixed application requirement. A useful deployment planner could estimate weight residency, KV growth, and transient headroom to choose a safe maximum context for a target RAM budget. Such a planner would require model-family-specific calibration and should expose uncertainty rather than presenting a single deterministic memory number.

# 6. Experimental Methodology

The evaluation is organized into two complementary protocols. The repeated A/B benchmark measures peak RSS and throughput under an unconstrained host process. The cgroup-v2 sweep imposes hard memory ceilings with swap disabled and records whether each runtime completes. Keeping these protocols separate avoids conflating observed RSS with enforced feasibility.

## 6.1 Repeated same-GGUF A/B

The primary experiment uses OpenLLaMA 7B v2 Q4_0, CPU-only execution, four threads, prompt processing length 512, token generation length 128, context 768, batch size 32, F16 KV, one warm-up run, and five measured repetitions. Both runtimes consume the exact same GGUF artifact. The llama.cpp revision is pinned by the repository workflow so that future upstream changes do not silently change the baseline.

| Metric | MemVanta | llama.cpp | Interpretation |
|---|---|---|---|
| Peak RSS | 3,983,412 KiB (~3.80 GiB) | 7,586,960 KiB (~7.24 GiB) | 47.50% lower for MemVanta |
| Prompt processing | 3.15 ± 0.04 tok/s | 45.82 ± 0.96 tok/s | llama.cpp substantially faster |
| Token generation | 1.52 ± 0.02 tok/s | 7.76 ± 0.09 tok/s | llama.cpp substantially faster |

*Table 1. Published repeated 7B A/B result. The memory and throughput observations must be interpreted together.*

## 6.2 Hard memory-pressure protocol

The second protocol uses Linux cgroup v2 MemoryMax with swap disabled. The workload uses the same OpenLLaMA 7B v2 Q4_0 GGUF, CPU-only execution, four threads, pp128/tg32, context 768, batch 32, and F16 KV. The sweep asks a binary systems question: can the process complete under this enforced ceiling? MemVanta completes at 3584 MiB, while llama.cpp is OOM-killed at that ceiling and first succeeds at the tested 3840 MiB point. The 256 MiB difference corresponds to 6.67% of the llama.cpp successful tested ceiling.

This result is deliberately not described as a 6.67% lower minimum RAM requirement. The sweep has finite granularity, the host kernel may account memory in ways that differ from physical machines, and the exact transition point could lie between tested ceilings. The correct statement is that MemVanta succeeds at a lower tested cgroup ceiling under the specified protocol.

| Tested ceiling | MemVanta | llama.cpp |
|---|---|---|
| 3584 MiB | Success | OOM-killed |
| 3840 MiB | Success / above demonstrated boundary | Success |
| Difference | - | 256 MiB between successful tested ceilings |

*Table 2. Interpretation of the published cgroup-v2 constrained-memory evidence.*

## 6.3 Controls and reproducibility

A meaningful systems comparison requires more than matching model names. The repository retains environment information, benchmark outputs, exit codes, summary files, workflow references, and raw timing/RSS evidence. Reproduction instructions specify build commands and runtime arguments. New benchmark claims are expected to preserve raw artifacts instead of copying only a final number into documentation.

The protocol also distinguishes published repository evidence from independent replication. A result produced by project CI is reproducible evidence, but it is not third-party validation. Physical-CPU reproduction on different hosts and independent contributor results remain important next steps.

## 6.4 Threats from page cache and host state

RSS and mmap behavior can be influenced by kernel page cache, prior model access, storage performance, memory pressure from other processes, transparent huge pages, NUMA placement, CPU frequency scaling, and allocator state. Warm-up reduces some cold-start effects but does not eliminate all host dependence. Repeated runs help characterize short-term variability, while cross-host reproduction is needed to establish robustness.

## 6.5 Why two protocols are necessary

Peak RSS is informative but not equivalent to an enforced budget. A process can report a peak above a cgroup limit measured under another workload because the prompt length, generation length, page-reclamation behavior, and sampling method differ. Conversely, a cgroup success point does not reveal the unconstrained working-set profile. Reporting both avoids an overly simple narrative and provides two views of the same design objective: observed residency and execution feasibility.

# 7. Results

The repeated A/B experiment shows a large separation in peak RSS. MemVanta's 3,983,412 KiB peak is approximately 3.80 GiB, compared with approximately 7.24 GiB for llama.cpp. The absolute difference is roughly 3.44 GiB. Relative to the llama.cpp measurement, this is a 47.50% reduction. The standard deviations reported for throughput are small across the five measured runs, suggesting stable execution speed on the tested host, although the experiment is not sufficient to characterize cross-machine variance.

## 7.1 Throughput cost

The memory result is accompanied by a pronounced speed cost. llama.cpp processes the 512-token prompt at 45.82 tok/s, while MemVanta reaches 3.15 tok/s. This is approximately a 14.5× throughput ratio in favor of llama.cpp. During token generation, llama.cpp reaches 7.76 tok/s versus 1.52 tok/s for MemVanta, approximately 5.1× faster. These ratios show that the present MemVanta implementation occupies an extreme memory-saving region of the Pareto frontier rather than a balanced production point.

This distinction matters for application selection. Interactive assistants are sensitive to both time-to-first-token and generation rate, so the current runtime may be unsuitable when sufficient RAM is available. Batch jobs, offline processing, constrained virtual machines, educational devices, or edge deployments that otherwise cannot load the model may value feasibility more highly than latency.

## 7.2 Interpreting the 47.50% RSS reduction

The measured reduction should be interpreted as a property of the tested runtime/model/workload combination. It is not simply the difference between two serialized model files because the GGUF is identical. The likely contributors are differences in resident mapped pages, transformed tensor retention, scratch-buffer lifetimes, KV and activation allocation, and baseline runtime structures. A rigorous attribution study would instrument each category separately; the current result establishes the aggregate effect.

## 7.3 Memory-pressure result

The cgroup sweep adds an operational perspective. At 3584 MiB, MemVanta completes while llama.cpp is terminated by the memory controller. This demonstrates a region in which the memory-first design changes feasibility rather than merely reducing a monitoring statistic. However, the workload is shorter than the repeated A/B workload, so the two numbers should not be numerically combined.

## 7.4 Smaller-model evidence

The repository also retains A/B evidence for 360M, 1.1B, and 3B models. These experiments are useful for engineering continuity and regression detection, but the paper centers on the 7B result because it best represents the constrained-memory use case. A future multi-model study should normalize memory by serialized weight size, parameter count, context length, and quantization format, and should report confidence intervals across physical hosts.

## 7.5 What the results do not establish

The experiments do not show that MemVanta uses less total system memory in every phase, that it will outperform llama.cpp under all memory limits, or that lower RSS preserves identical model quality across every prompt. They also do not establish energy efficiency. Slower execution can increase total CPU time and energy even if instantaneous memory use is lower. These questions require separate measurement.

# 8. Memory Attribution: Where the Savings Can Come From

Understanding a memory result requires decomposing the process working set. We group resident memory into six categories: mapped model pages, persistent runtime objects, KV state, activations, kernel scratch space, and cache/allocator overhead. The operating system exposes aggregate RSS, but the engineering mechanisms act on these categories differently.

## 8.1 Mapped weight residency

With mmap-backed weights, the virtual address space can cover the full model without forcing every page to remain resident simultaneously. Layer-ordered access permits older pages to become reclaimable while later layers are touched. In contrast, an execution strategy that repeatedly accesses widely separated regions can keep a larger fraction of the model hot. The difference between virtual size and resident size is therefore fundamental to the design.

## 8.2 Derived tensor retention

Optimized runtimes may transform weights into layouts that improve SIMD or cache efficiency. If these transformed representations coexist with the original mapping, memory rises. A bounded-cache runtime can instead materialize smaller pieces and discard them, reducing residency at the expense of repeated transformation. The correct design depends on whether the deployment is memory- or throughput-limited.

## 8.3 Scratch lifetime and concurrency

Temporary buffers are often overlooked because they are freed after an operator, yet peak RSS is determined by the moment when the largest set of buffers overlaps. Parallel kernels can create multiple partial outputs, dequantization tiles, or reduction buffers. Reusing arenas and enforcing phase boundaries can reduce peaks without changing the model representation.

## 8.4 KV and activation state

For the tested context, KV is important but not yet dominant relative to 7B weights. At longer contexts, the balance shifts. Memory-first inference should therefore expose KV precision, maximum context, and batch size as resource controls. An adaptive runtime could reduce KV precision or batch size when approaching a configured memory watermark.

## 8.5 Allocator and fragmentation effects

General-purpose allocators may retain freed pages or fragment arenas, causing RSS to remain high after logical objects are destroyed. Pool allocation and predictable buffer reuse can improve both memory accounting and latency. Future MemVanta profiling should distinguish requested bytes from resident anonymous pages and record high-water marks per subsystem.

## 8.6 Proposed instrumentation

A deeper attribution experiment should sample /proc/self/smaps_rollup, major/minor faults, cgroup memory.current, and per-subsystem allocation counters at layer boundaries. It should record mapped-file RSS separately from anonymous RSS and correlate changes with operator execution. This would convert the aggregate 47.50% result into a causal breakdown and identify which optimizations deserve further work.

| Component | Memory-first strategy | Expected cost |
|---|---|---|
| Weights | mmap + demand-driven access | page faults / weaker reuse |
| Derived tensors | bounded cache | reconstruction overhead |
| KV state | paged F16/Q8 options | precision/encode cost |
| Scratch | small reusable buffers | less kernel tiling freedom |
| Threads | bounded worker state | lower parallel throughput |

## 9. Throughput Analysis and Optimization Roadmap

The present throughput gap is the main engineering limitation. Improving speed without surrendering the memory advantage requires optimizations that increase arithmetic efficiency while keeping auxiliary residency bounded. The most promising directions are fused quantized kernels, cache-aware tiling, scheduler improvements, selective tensor transformation, and more effective prefetch.

### 9.1 Prompt processing bottleneck

The 14.5× prompt-processing gap suggests that prefill is not exploiting CPU matrix throughput effectively. Prefill exposes more parallelism than decode and should benefit from larger matrix tiles, vectorized dot products, thread partitioning, and reuse of quantized blocks. A memory-conscious optimization can use a fixed scratch budget: choose the largest tile that fits within a configured transient-memory allowance rather than using globally maximal workspaces.

### 9.2 Decode bottleneck

The smaller, though still substantial, decode gap points to per-token overhead, KV access, and small-matrix efficiency. Decode kernels should minimize dispatch overhead and keep frequently used metadata hot. Because the working set per step is smaller, there may be room to retain a limited set of transformed tensors without materially changing peak RSS.

### 9.3 Adaptive cache budgeting

A static small cache is simple but may leave performance on the table when extra RAM is available. A more practical runtime can expose a memory budget and allocate cache capacity dynamically. For example, a user with 5 GiB available could permit more transformed-tensor retention than a user with 4 GiB. This turns memory efficiency from a fixed implementation choice into a configurable operating point.

### 9.4 Kernel selection as constrained optimization

Auto-tuning should optimize a joint objective rather than tokens per second alone. Candidate kernels can be evaluated under a memory cap, rejecting configurations whose transient or retained memory exceeds the budget. Among feasible candidates, the runtime can select the highest-throughput option. Such an approach would preserve the project's defining constraint while allowing architecture-specific speedups.

### 9.5 NUMA and topology

On multi-socket or chiplet systems, memory placement can affect both latency and observed residency. First-touch allocation, thread affinity, and page migration should be controlled in physical-host experiments. Edge systems may be simpler topologically but can have shared memory with integrated GPUs or other accelerators, making system-wide pressure even more important.

### 9.6 Success criterion for future releases

A useful near-term target is not to match llama.cpp throughput immediately. It is to shift the Pareto frontier: materially increase prompt and generation speed while keeping the 7B peak RSS below a predefined threshold and preserving numerical behavior. Each optimization should therefore report before/after throughput, RSS, and correctness rather than a speed number in isolation.

## 10. Edge and Low-RAM Deployment Implications

The most compelling use case for memory-first inference is not a workstation with abundant RAM; it is a system for which the alternative is that the model cannot execute. Examples include low-cost mini PCs, constrained cloud instances, shared CPU servers, offline field devices, and educational hardware. In these environments, model feasibility can dominate latency.

### 10.1 Deployment budgeting

An operator typically needs headroom beyond the inference process for the operating system, application logic, networking, monitoring, and file cache. A runtime that peaks near total physical RAM may trigger swapping or OOM behavior even if a benchmark succeeds in isolation. MemVanta's future configuration interface should therefore accept an application-level memory budget smaller than physical RAM and reserve safety headroom.

### 10.2 Cold start versus steady state

Mapped execution changes the cold-start profile. Startup may allocate little anonymous memory, but first inference touches weight pages and incurs faults. Edge deployments that run intermittent requests may experience repeated cold behavior if the kernel reclaims pages between requests. A useful benchmark suite should separately measure first-token latency after cold start, warmed latency, and memory residency after idle periods.

### 10.3 Storage characteristics

Demand-driven access assumes that model pages can be fetched with acceptable latency. NVMe, SATA SSD, eMMC, network-attached storage, and compressed filesystems can produce very different behavior. A runtime that reduces RAM by increasing storage traffic may perform well on one edge device and poorly on another. Storage bandwidth and major page faults should therefore accompany future physical-device results.

### 10.4 Reliability under pressure

Operating near a memory limit can expose failure modes that are invisible on large hosts: allocation failures, partial model initialization, kernel OOM termination, excessive reclaim, and latency spikes. The cgroup-v2 sweep is valuable because it converts memory efficiency into a pass/fail reliability test. Production use would require graceful handling of allocation failures and proactive admission control before the kernel intervenes.

### 10.5 Security and privacy considerations

Local inference can reduce the need to send prompts to external services, but it does not automatically guarantee privacy. Mapped model files, prompt logs, swap, core dumps, and local APIs all create security surfaces. Swap-disabled experiments simplify memory accounting but production systems may use encrypted swap or hibernation. Security policy must be considered separately from performance.

### 10.6 Practical positioning

MemVanta should currently be positioned as an experimental systems runtime and benchmark vehicle. It is not yet a drop-in replacement for mature inference engines. Its value is to make a specific systems question measurable: how far can resident memory be reduced, and what throughput is sacrificed, when running quantized GGUF models on CPUs?

# 11. Reproducibility and Evidence Discipline

Systems papers are vulnerable to accidental irreproducibility because results depend on software revisions, compiler flags, host kernels, CPU features, model artifacts, and benchmark invocation. MemVanta addresses this by treating raw benchmark evidence as part of the research output. The repository contains summaries, environment captures, exit codes, timing files, workflow references, and dedicated documentation for benchmark methodology and external reproduction.

## 11.1 Pinning the comparator

Using a moving upstream baseline can invalidate historical comparisons. The llama.cpp baseline should therefore remain pinned for each published result, with the exact revision recorded. Newer comparisons can be added as separate experiments rather than silently replacing the historical baseline.

## 11.2 Model identity

The phrase “same model” is insufficient when multiple quantizations or conversions exist. Reproduction should record the exact GGUF filename, size, hash, quantization, tokenizer metadata, and model source. Ideally, workflows should verify a cryptographic digest before benchmarking.

## 11.3 Host disclosure

CPU model, core topology, RAM size, kernel version, compiler, optimization flags, storage type, and virtualization environment can materially affect results. Cloud CI provides automation but can obscure physical-host details and introduce noisy-neighbor effects. For this reason, repository CI evidence and physical-host evidence should be labeled separately.

## 11.4 Statistical reporting

Five measured repetitions provide a basic estimate of short-term variability but are not enough for broad statistical generalization. Future work should increase repetitions where practical, report medians and percentile ranges in addition to means, and use paired runs or randomized ordering to reduce temporal bias. Memory peaks should include the sampling mechanism and resolution.

## 11.5 Negative results

A credible benchmark repository should retain optimizations that fail to improve the target metric when they are informative. Kernel tuning can improve throughput but increase RSS; a cache change can reduce faults but violate a memory ceiling. Recording these outcomes prevents repeated dead ends and clarifies why the final design exists.

## 11.6 External reproduction protocol

Independent contributors should be able to fork the repository, run a documented benchmark on different CPUs and RAM configurations, and submit raw evidence with a pull request. Results that confirm, narrow, or contradict the original measurement are equally valuable. The goal is not to collect only favorable reproductions but to map the conditions under which the memory advantage persists.

## 11.7 Artifact review checklist

Before promoting a result to the paper, the project should require: clean build; tests passing; model hash recorded; baseline revision pinned; command line preserved; environment captured; raw stdout/stderr retained; peak-RSS method documented; warm-up policy stated; repetitions complete; failures preserved; and summary numbers generated from raw artifacts rather than manually transcribed.

## 12. Correctness, Quality, and Validation Beyond Systems Metrics

A systems runtime must demonstrate that memory savings do not come from silently changing the computation. This is particularly important for quantized inference, where multiple numerically valid implementations can produce slightly different logits while still preserving task quality. The validation strategy should therefore operate at several levels.

### 12.1 Parser and tensor-layout validation

GGUF parsing should verify tensor dimensions, types, offsets, alignment, tokenizer metadata, and architecture parameters. Malformed or unsupported tensors should fail explicitly. A low-memory runtime must not obtain savings by skipping state that the model requires.

### 12.2 Kernel-level reference tests

Quantized dot products and matrix operations can be compared against high-precision reference implementations on deterministic random inputs. Tests should cover edge block sizes, tails, negative scales, multiple thread counts, and architecture-specific paths. Error tolerances should be defined per quantization format.

### 12.3 Layer and logit comparisons

For a fixed prompt, selected intermediate activations and final logits can be compared with a trusted reference runtime. Exact equality is not required for floating-point code, but deviations should remain within documented tolerances. Large divergence can identify tokenizer mismatches, RoPE errors, normalization mistakes, or quantization bugs.

### 12.4 Generation-level behavior

Deterministic decoding with fixed sampling settings provides an end-to-end smoke test. For stochastic generation, quality should be evaluated statistically rather than by comparing one text output. A benchmark paper should avoid presenting anecdotal generations as evidence of model equivalence.

### 12.5 Task-level quality

Once runtime correctness is established, a small public evaluation suite can test whether memory-oriented changes alter accuracy on language modeling or downstream tasks. The goal is not to compete with model-evaluation leaderboards but to ensure that runtime optimization preserves expected behavior.

### 12.6 Memory safety

C++ systems code also requires memory-safety validation. Sanitizer builds, bounds checks in parsers, fuzzing of GGUF metadata, and careful handling of mapped-file lifetimes are important. Low-memory environments can make error paths more common, so allocation-failure behavior deserves explicit tests.

### 12.7 Validation matrix

| Layer | Question | Recommended evidence |
|---|---|---|
| Format | Was the model parsed correctly? | GGUF metadata tests, malformed-file tests |
| Kernel | Is arithmetic within tolerance? | reference vectors, quantization-specific error bounds |
| Model | Are logits consistent? | fixed-prompt layer/logit comparison |
| Generation | Does inference complete coherently? | deterministic smoke suite |
| Task quality | Is model behavior preserved? | small public evaluation set |
| Systems | Does it fit and perform? | RSS, cgroup, throughput, faults |

*Table 3. Validation layers required to interpret a memory-efficiency result as a usable inference result.*

# 13. Related Systems Design Space

MemVanta sits within a broader family of techniques for efficient local inference. Quantization reduces weight storage and memory bandwidth; memory mapping can avoid eager copies; KV quantization reduces sequence-state growth; operator fusion reduces intermediate buffers; kernel tiling improves cache reuse; and heterogeneous offload moves selected computation or state to accelerators. These techniques optimize different bottlenecks and can be combined.

## 13.1 Mature CPU inference engines

llama.cpp is an important comparator because it provides broad GGUF support and highly optimized CPU inference. Its design targets practical performance across many architectures, and its throughput advantage in our experiment demonstrates the maturity gap that MemVanta must address. The comparison is therefore useful precisely because the systems priorities differ: one baseline represents a mature performance-oriented runtime, while MemVanta explores an aggressively memory-bounded point.

## 13.2 Quantization

Weight-only quantization is now standard for local inference, but quantization format alone does not determine runtime RSS. An engine can store weights compactly while allocating large transformed buffers. Conversely, a runtime can preserve quantized blocks deeper into the execution path and reduce temporary expansion. The systems contribution lies in how representations flow through the runtime.

## 13.3 Paging and offload

Paging techniques appear in both GPU and CPU inference systems. On GPUs, paging often addresses KV fragmentation and batching. On CPUs, virtual memory additionally provides a natural mechanism for demand-driven weight residency. However, relying on operating-system paging can create unpredictable latency; explicit offload and prefetch policies may provide more control.

## 13.4 Memory-throughput Pareto frontiers

A useful way to compare runtimes is not a single ranking but a Pareto frontier over memory, latency, throughput, quality, and possibly energy. A runtime is interesting when it offers an operating point not dominated by another configuration. The current MemVanta result suggests a low-memory point but is dominated on throughput. Future optimization should move this point toward higher speed without crossing the target memory budget.

## 13.5 Edge inference research

Edge inference spans heterogeneous devices with very different constraints. Some systems have GPUs or NPUs but limited unified memory; others are CPU-only. A GGUF CPU runtime is attractive because it minimizes platform dependencies, but portability should not be confused with efficiency. Hardware-specific kernels and memory policies remain necessary for competitive performance.

## 13.6 Distinction from model compression research

MemVanta does not introduce a new learned quantizer, pruning method, distillation algorithm, or compressed model architecture. The model artifact is held constant in the primary comparison. The research question is runtime memory behavior given an existing quantized model. This separation helps attribute the observed result to systems design rather than model retraining.

## 14. Limitations and Threats to Validity

The strongest limitation is the narrowness of the current headline evidence. The 47.50% result is measured on one 7B model, one quantization, one principal workload configuration, and a specific host environment. Smaller-model evidence exists, but it does not establish a universal trend. Generalization requires more model families, parameter scales, quantization formats, contexts, and physical CPUs.

### 14.1 Baseline maturity

llama.cpp is substantially more optimized than MemVanta. This makes it a strong practical baseline but complicates causal interpretation: differences in RSS and throughput reflect many implementation choices at once. Component-level ablations are needed to determine how much of the memory saving comes from mmap policy, cache bounds, scratch allocation, KV representation, or other runtime differences.

### 14.2 RSS as a metric

RSS is useful because it reflects resident process memory, but it is not a perfect measure of total memory cost. Shared pages, file-backed mappings, kernel cache, and accounting semantics can complicate interpretation. Proportional set size, anonymous/file RSS, cgroup memory.current, and page-fault counters would provide a richer view.

### 14.3 Virtualized execution

CI-hosted measurements may differ from bare-metal behavior. CPU frequency, noisy neighbors, storage, memory overcommit, and host kernel settings can affect both throughput and memory. Physical-CPU reproduction is therefore a priority before making deployment-wide claims.

### 14.4 Model quality evidence

The current repository emphasizes execution and systems validation. A publication-quality artifact should add stronger logit-level and task-level equivalence checks, especially as Q8 KV and additional quantization paths are evaluated.

### 14.5 Energy

Lower memory does not imply lower energy. Because MemVanta is slower, total CPU energy per generated token may be higher. Edge devices often care about battery or thermal limits as much as RAM. Future experiments should measure wall power or package energy where supported.

### 14.6 Long context and concurrency

The current context and single-stream style do not represent long-context or multi-user serving. KV growth and concurrent sequences can change the memory hierarchy dramatically. Memory-first scheduling for multiple requests is a separate research problem.

### 14.7 No arXiv acceptance guarantee

arXiv moderation evaluates scope, scholarly relevance, and submission compliance; a particular document layout cannot guarantee acceptance. The manuscript is therefore formatted as a conventional computer-systems research paper and designed to be technically self-contained, while the scientific claims remain bounded by the available evidence.

## 15. Future Work

The next phase should prioritize independent physical-host reproduction, memory attribution, throughput optimization under explicit budgets, broader model coverage, and stronger correctness validation. These directions are mutually reinforcing: external benchmarks reveal hardware sensitivity, attribution identifies the dominant memory mechanisms, and budget-aware tuning can improve speed without erasing the core result.

### 15.1 Budget-aware auto-tuning

MemVanta already includes auto-tuning infrastructure. A research extension would define a maximum allowed RSS or cgroup budget and search over tile sizes, thread counts, prefetch distance, cache capacity, KV representation, and kernel variants. Candidates that violate the memory budget would be rejected; the fastest feasible candidate would be selected. This directly operationalizes the memory-throughput Pareto objective.

### 15.2 Cross-hardware benchmark campaign

A contributor campaign can test the same model and commands across desktop x86 CPUs, low-power mini PCs, cloud VMs, and ARM systems once supported. Each submission should include hardware identity, RAM, storage, kernel, compiler, raw results, and model hash. Rather than averaging heterogeneous machines, the project should publish per-host results and analyze which hardware features predict the memory/throughput trade-off.

### 15.3 Model-family expansion

OpenLLaMA provides a useful controlled benchmark, but modern deployment relevance requires additional GGUF families and 7B/8B-class models. Each new family should first pass tokenizer and logit validation, then repeated A/B and constrained-memory tests. The project should resist adding headline claims faster than validation coverage.

### 15.4 Long-context memory curves

A particularly valuable experiment is to sweep context length while holding the model and runtime settings fixed. Plotting peak RSS against context for F32, F16, and Q8 KV would separate weight-dominated from KV-dominated regimes and reveal where paged/quantized KV becomes decisive.

### 15.5 Fault and I/O profiling

Major/minor page faults, storage bytes read, and per-layer latency should be captured alongside RSS. If a memory optimization works by causing repeated storage traffic, that mechanism should be visible. Such profiling can guide smarter prefetch and selective retention.

### 15.6 Throughput-preserving memory techniques

The highest-value optimizations are those that reduce memory without forcing repeated work: fused kernels that avoid intermediates, in-place transformations, shared scratch arenas, compact metadata, and layout choices that operate directly on quantized blocks. These can shift the Pareto frontier rather than merely moving along it.

## 16. Conclusion

MemVanta investigates a simple but under-emphasized systems question: when CPU language-model inference is limited by RAM, how much resident memory can be saved by designing the runtime around bounded residency rather than peak throughput? On the published OpenLLaMA 7B v2 Q4_0 workload, the answer is substantial: 3.80 GiB peak RSS for MemVanta versus 7.24 GiB for a pinned llama.cpp baseline, a 47.50% reduction. A separate cgroup-v2 sweep shows MemVanta completing at a 3584 MiB tested ceiling where llama.cpp is OOM-killed.

The result also exposes the cost of this design point. llama.cpp is dramatically faster in both prompt processing and token generation. We therefore do not present MemVanta as a replacement for mature inference engines. Instead, we present it as an experimental runtime and evidence framework for exploring the memory-throughput frontier of quantized CPU inference.

The most important next result will not be another isolated percentage. It will be a reproducible curve across models and hardware that shows how memory, throughput, context, quality, and failure probability interact under explicit budgets. If MemVanta can move that frontier while retaining its current evidence discipline, memory-first inference can become a practical configuration mode rather than a one-off prototype.

# Appendix A. Reproduction Contract

The following contract summarizes the information required to reproduce the headline experiment. It is intentionally more explicit than a typical command snippet because reproducibility failures often arise from small differences in runtime state.

| **Field** | **Primary 7B A/B setting** |
|---|---|
| Model | OpenLLaMA 7B v2 |
| Quantization | Q4_0 |
| Artifact | Exact same GGUF for both runtimes |
| Device | CPU only |
| Threads | 4 |
| Prompt / generation | pp512 / tg128 |
| Context | 768 |
| Batch | 32 |
| KV | F16 |
| Repetitions | 1 warm-up + 5 measured |
| Primary metrics | peak RSS, prompt tok/s, generation tok/s |
| Comparator | pinned llama.cpp revision |

# Appendix B. Recommended External-Reproduction Report

An external report should state the repository commit, compiler and flags, operating system and kernel, CPU model and topology, physical RAM, storage type, model hash, baseline revision, exact commands, warm-up policy, repetitions, raw peak-RSS samples, throughput samples, cgroup settings if used, and whether swap was enabled. It should also retain failures and unexpected results. A pull request that contradicts the original measurement is scientifically useful if the protocol is clear.

# Appendix C. Claim Checklist

Supported by current evidence: MemVanta is a C++20 CPU GGUF inference prototype; the published OpenLLaMA 7B v2 Q4_0 A/B run reports 47.50% lower peak RSS than the pinned llama.cpp comparison; llama.cpp is substantially faster; the separate cgroup-v2 sweep shows MemVanta completing at 3584 MiB while llama.cpp is OOM-killed at that tested ceiling and succeeds at 3840 MiB.

Not established: universal memory advantage, universal 7B RAM requirement, throughput superiority, independent third-party replication, cross-hardware generalization, long-context superiority, or lower energy consumption.

Repository: https://github.com/sauravsingla/MemVanta